\documentclass[aps,showpacs,twocolumn,preprintnumbers,amsmath,amssymb,prl]{revtex4-1}

\usepackage{graphicx}
\usepackage{dcolumn}
\usepackage{bm}
\usepackage{color}

\begin{document}

\title{Resonant radiation shed by dispersive shock waves}

\author{Matteo Conforti$^1$, Fabio Baronio$^1$, Stefano Trillo$^2$}

\affiliation{$^1$CNISM, Dipartimento di Ingegneria dell'Informazione, Universit\`a di Brescia, Via Branze 38, 25123 Brescia, Italy\\
$^2$ Dipartimento di Ingegneria, Universit\`a di Ferrara, Via Saragat 1, 44122 Ferrara, Italy}
\date{\today}

\begin{abstract}
We show that dispersive shock waves resulting from the nonlinearity overbalancing a weak leading-order dispersion can emit resonant radiation owing to higher-order dispersive contributions. We analyze such phenomenon for the defocusing nonlinear Schr\"odinger equation, giving criteria for calculating the radiated frequency based on the estimate of the shock velocity, revealing also a diversity of possible scenarios depending on the order and magnitude of the dispersive corrections.
\end{abstract}

                              
\maketitle
Dispersive shock waves (DSWs) are expanding regions filled with fast oscillations that stem from the dispersive regularization of classical shock waves (SWs). Originally introduced in collisionless plasmas \cite{collisionless} and water waves \cite{water1}, it is only recently that DSWs have been the focus of intense multidisciplinary efforts that have established their universal role in atom condensates \cite{BEC,Hoefer06}, light pulse (temporal) \cite{temporal} and beam (spatial) \cite{spatial} propagation, oceanography \cite{water}, quantum liquids \cite{quantum}, electron beams \cite{electrons}, magma flow \cite{magma}, and granular \cite{granular} or disordered materials \cite{disorder}. The dynamics of DSW is understood in terms of {\em weakly dispersive formulation} of integrable models (and their defomations) such as the Korteweg De Vries  \cite{collisionless,electrons}, the Benjamin-Ono \cite{quantum,BO} or the {\em defocusing} nonlinear Schr\"odinger equation (dNLSE)  \cite{Hoefer06,temporal,spatial,dNLSE}. However, since the leading order dispersion of such models must be extremely weak for the phenomenon to take place, one is naturally led to wonder about the effects of higher-order dispersion (HOD). The aim of this Letter is to show that HOD corrections lead DSWs to emit resonant radiation (RR) due to a specific phase-matching with linear waves.\\ 
\indent The emission of RR, usually considered for solitons \cite{RRoptics,AK95,RRdark,RRothers,DudleyRMP06,SGRMP10,SChydro,Stark11,Joly11,Colman12,Erkintalo12},
is a relatively well understood phenomenon in connection with studies of supercontinuum generation driven by perturbed solitons of the {\em focusing} NLSE (fNLSE) \cite{DudleyRMP06,SGRMP10,SChydro}, and the emergence of novel regimes \cite{Stark11,Joly11,Erkintalo12,Colman12,Bache10,rubino12}. Viceversa, the problem of RR from SWs was overlooked. Here we show that perturbed DSWs emit RR, owing to a strong spectral broadening that accompanies wave-breaking and acts as a seed for linear waves that are resonantly amplified thanks to a well defined velocity of the shock front. While we expect this mechanism to be universal for several DSW-bearing models when HOD corrections become effective, we formulate our approach for temporal pulse propagation ruled by the dNLSE \cite{spatialHOD}, where our results are important in view of generating a different type of supercontinuum pumped in the normal dispersion regime \cite{SC12}. They have also immediate impact to unveil the underlying mechanism of recent observations of RR produced by non-soliton pulses \cite{Webb13}. We start from the dNLSE in semiclassical form \cite{spatial}, with dispersion at all orders (sum over $n$ implicitly assumes $n \ge2$)
\begin{eqnarray} \label{wdnls}
&i \varepsilon \partial_z \psi + d(\partial_t) \psi + |\psi|^2 \psi = 0,\\
&d(\partial_t) \equiv \sum_{n}  \frac{\beta_n}{n!} \varepsilon^n\,( i\partial_t )^n
=- \frac{\varepsilon^2}{2} \partial^2_t - i  \frac{\beta_3 \varepsilon^3}{6} \partial^3_t + \frac{\beta_4 \varepsilon^4}{24} \partial^4_t + \ldots \nonumber
\end{eqnarray}
where the link with real-world distance and retarded time (in capital) is $Z=z\sqrt{L_{nl} L_d}$, $T=t T_0$, where $L_{nl}=(\gamma P)^{-1}$ and $L_d=T_0^2/\partial_\omega^2 k$ are the nonlinear and dispersive length, respectively, associated with input pulse width $T_0$ and peak power $P$ ($\gamma$ is the nonlinear coefficient). The dispersive operator $d(\partial_t)$ have terms which are weighted, without loss of generality,  by growing powers of the small parameter $\varepsilon=\sqrt{L_{nl}/L_d} \ll 1$ and coefficients $\beta_n= \partial_\omega^n k/\sqrt{(L_{nl})^{n-2} (\partial_\omega^2 k)^{n}}$ (note that $\beta_2=1$), $\partial_\omega^n k$ being $n$-order real-world dispersion arising from usual Taylor expansion of $k(\omega)$ (further details on supplemental material \cite{SM}).
We assume an input pump $\psi_0=\psi(t,z=0)$ with central frequency $\omega_p=0$ \cite{nota_wp},
and denote as $V_s$ the ``velocity'' of the SW produced by $\psi_0$ via wave-breaking (here, $V_s=dt/dz$ is the reciprocal of the velocity as usually defined for soliton RR \cite{AK95}), and as $\tilde{d}(i\omega)$ the Fourier transform of $d(\partial_t)$. Linear waves $\exp(i k(\omega) z - i \omega t)$ are resonantly amplified when their wavenumber in the SW moving frame, which reads as $k(\omega)=\frac{1}{\varepsilon} \left[\tilde{d}(i\omega)-V_s (\varepsilon \omega) \right]$
equals the pump wavenumber $k_p=k(\omega_p=0)=0$. Denoting also as $k_{nl}$ the difference between the nonlinear contributions to the pump and RR wavenumber \cite{notaRR}, respectively,
the radiation is resonantly amplified at frequency $\omega=\omega_{RR}$ which solves the equation
\begin{equation}\label{omRR}
\sum_n \frac{\beta_n}{n!} (\varepsilon \omega)^n -  V_s (\varepsilon \omega)  = \varepsilon k_{nl}.
\end{equation}
At variance with solitons of the fNLSE where $V_s(\omega_p=0)=0$ \cite{AK95,SGRMP10}, DSWs possess non-zero velocity $V_s$, which must be carefully evaluated, having great impact on the determination of $\omega_{RR}$.

The process of wave-breaking ruled by Eq. (\ref{wdnls}) can be described by applying the Madelung transformation $\psi=\sqrt{\rho} \exp \left( i S/\varepsilon \right)$.
At leading-order in $\varepsilon$, we obtain  a quasi-linear hydrodynamic reduction, with $\rho=|\psi|^2$ and $u=-S_t$ (chirp) equivalent density and velocity of the flow,
which can be further cast in the form
\begin{eqnarray}
&\partial_z \rho  + \partial_t \left[ \sum_{n}  \frac{\beta_n}{(n-1)!} (\rho u^{n-1}) \right] = 0, \label{cons1} \\
&\partial_z (\rho u)  + \partial_t \left[\sum_{n}  \frac{\beta_n}{(n-1)!} \rho u^{n} + \frac{1}{2} \rho^2 \right] =0  \label{cons2},
\end{eqnarray}
of a conservation law $\partial_z {\bf q} + \partial_t {\bf f}({\bf q})=0$ for mass and momentum, with ${\bf q}=(\rho, \rho u)$.
This system can be diagonalized to yield
$\partial_z r^{\pm} + V^{\pm} \partial_t r^{\pm} = 0$,
by introducing the eigenvelocities $V^\pm=\sum_{n}  \beta_n u^{n-1}/(n-1)! \pm \sqrt{\rho \sum_n \beta_n u^{n-2}/(n-2)! }$
and the Riemann invariants $r^{\pm} = u \pm 2 \frac{\sqrt{\rho}}{ \sqrt{\sum_n \beta_n u^{n-2}/(n-2)!} }$.

Equations (\ref{cons1}-\ref{cons2}), as far as HOD is such that they remain hyperbolic, admit weak solutions in the form of classical SWs, i.e. traveling discontinuity from left ($\rho_l, u_l$) to right ($\rho_r, u_r$) values, whose velocity $V_c$ can be found from the so-called Rankine-Hugoniot (RH) condition $V_c ({\bf q}_l-{\bf q}_r)=[{\bf f}({\bf q}_l)-{\bf f}({\bf q}_r)]$ \cite{Whitham74}. In the $2 \times 2$ case, the RH equations fix both $V_c$ and the admissible value of one of the parameters of the jump, e.g. $u_r$ given $\rho_{r}, \rho_{l}, u_l$.
For instance, when no HOD is effective (take $\beta_2=1$), an admissible right-going shock which satisfies the entropy condition $\rho_l>\rho_r$, can be obtained with 
\FL \begin{equation}\label{RH}
u_r =  u_l - (\rho_l-\rho_r)\sqrt{ \frac{\rho_r + \rho_l}{2\rho_l\rho_r} }; V_c= u_l  + \rho_r\sqrt{ \frac{\rho_r + \rho_l}{2\rho_l\rho_r}}.
\end{equation} 
This result can be generalized for HOD, thanks to Eqs. (\ref{cons1}-\ref{cons2}). For instance, if $\beta_3 \neq 0$, the SW velocity becomes 
\begin{eqnarray} \label{RHtod}
V_c=\frac{\beta_2(\rho_l u_l - \rho_r u_r) + \beta_3(\rho_l u_l^2 - \rho_r u_r^2)/2}{(\rho_l - \rho_r)},
\end{eqnarray}
where $u_r$ is obtained as the real root of the cubic equation $\beta_3 (u_l-u_r)^2(u_l+u_r)+ 2 \beta_2 (u_l-u_r)^2 = g(\rho_l,\rho_r)$,
where $g(\rho_l,\rho_r) \equiv (\rho_l-\rho_r)^2(\rho_r + \rho_l)/(\rho_l\rho_r)$ (see supplemental for more details \cite{SM}).\\
\begin{figure}[b]
{\includegraphics[width=4.25cm]{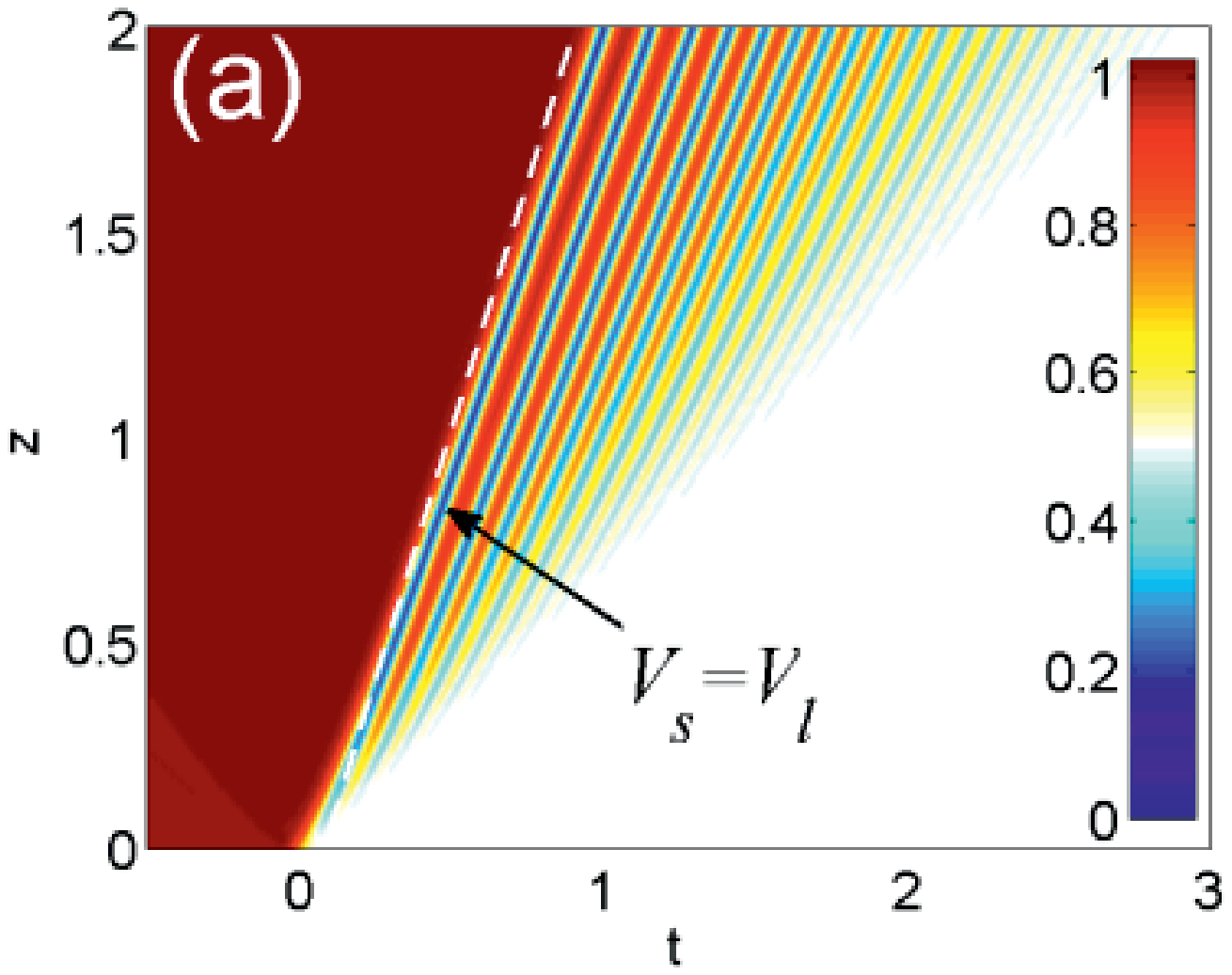}}
{\includegraphics[width=4.25cm]{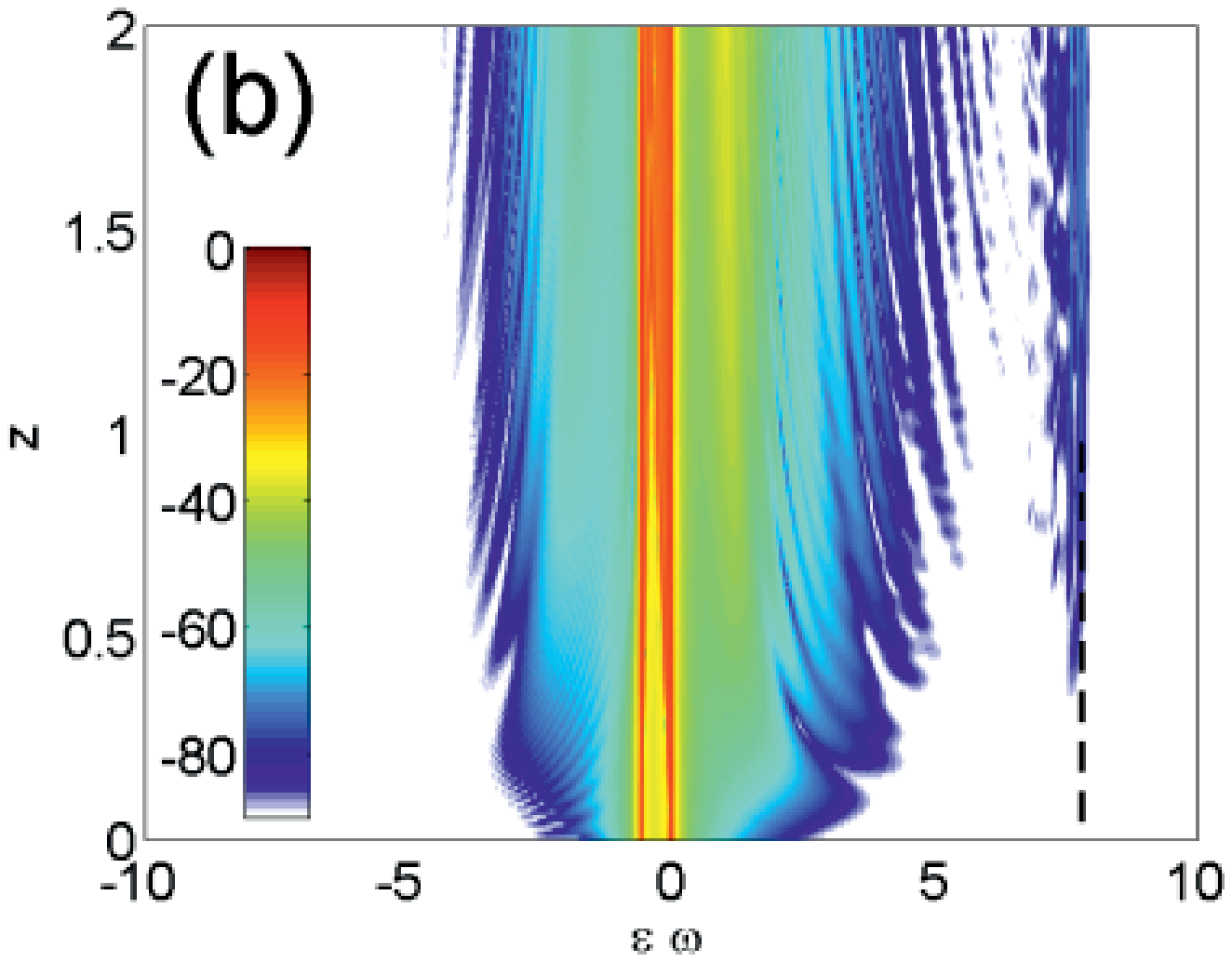}}
{\includegraphics[width=4.25cm]{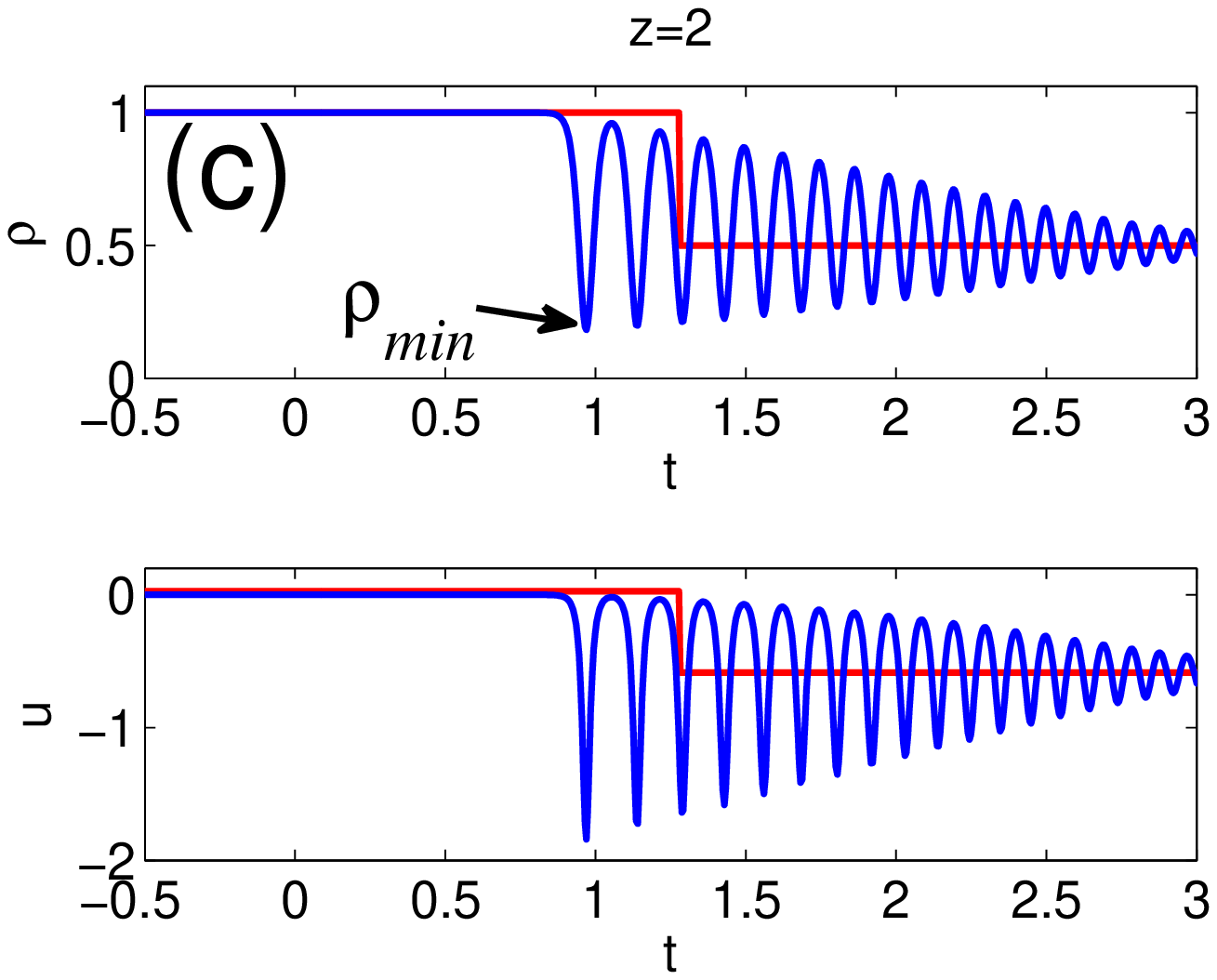}}
{\includegraphics[width=4.25cm]{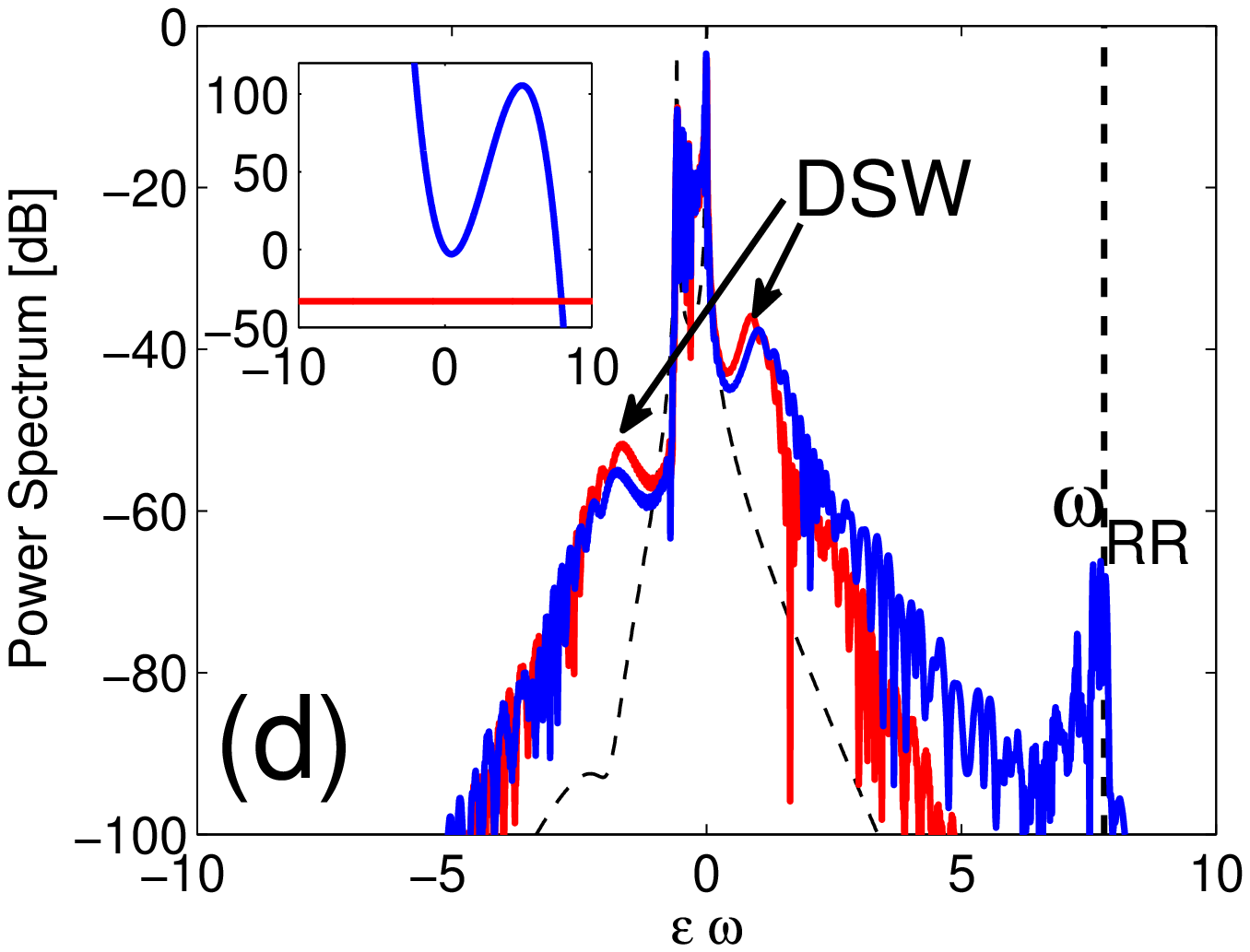}}
\caption{(Color online) Radiating DSW from dNLSE (\ref{wdnls}) with $\varepsilon=0.03$, input step $\rho_l, \rho_r=1,0.5$, 
and 3-HOD $\beta_3=-0.35$: (a) Color level plot of density $\rho(t,z)$ (the dashed line gives the DSW leading edge velocity $V_l$); 
(b) spectral evolution;
(c) snapshots of $\rho, u$ of unperturbed case $\beta_3=0$ (in red the corresponding classical SW); 
(d) output spectra with (blue) and without (red) 3-HOD (input dashed); inset: graphical solution of Eq. (\ref{omRR}).}
\label{fig1}
\end{figure}
Second-order dispersion, however, is known to regularize classical SWs by replacing the jump with an expanding fan filled with oscillations (i.e. a DSW) characterized
by leading $V_l$ and trailing $V_t$ edge velocities (with $V_l< V_c < V_t$), where the modulated periodic wave locally tends to a soliton and a linear wave, respectively \cite{Hoefer06}.
HOD induces this structure to radiate, also altering the dynamics of SW formation. 
Hereafter we specifically focus on the effect of two leading HOD, namely third-order (3-HOD) and fourth-order (4-HOD) dispersion.

In particular, when 3-HOD is effective we find a cross-over from a perturbative regime ($|\beta_3| \lesssim 0.5$) where the DSW leading edge turns out to be responsible for the RR, to a regime where the 3-HOD is strong enough ($|\beta_3| \sim 1$) to modify the shock formation, leading to enhanced RR produced by a traveling front which is approximated with a classical SW. To show this and verify that Eq. (\ref{omRR}) is able to predict the RR frequency in both regimes, we consider first a step initial value that allows us to calculate the velocity analytically, taking $\beta_3<0$ without loss of generality. Specifically, we consider the evolution of an initial jump from the state $\rho_l, u_l=0$ to the state $\rho_r, u_r=2(\sqrt{\rho_r} -\sqrt{\rho_l})$, which is such to maintain constant $r^-(z,t)$ upon evolution (only $r^+(z,t)$ varies). In this case, the modulated wavetrain produced upon evolution [see Fig. 1(a,c)] in the limit $\beta_3=0$ is described by a rarefaction wave of the Whitham modulation equations for the unperturbed dNLSE \cite{Hoefer06}. Following the approach of Ref. \cite{Hoefer06}, one can calculate the edge velocities of the fan. What is relevant for the RR is the leading-edge velocity, which we find to be $V_l=\sqrt{\rho_l} + u_r=2\sqrt{\rho_r} -\sqrt{\rho_l}$. 
Given a gray soliton on unchirped background $A$, $\psi=A [w \tanh (\theta) + i v ] \exp(i A^2 z/\varepsilon)$, $\theta=\frac{w}{\varepsilon} (t- A v z)$, $w^2=1-v^2$, $V_l$ turns out to coincide with the soliton velocity $V_{sol}=Av=\sqrt{\rho_{min}}$, with natural position $A=\sqrt{\rho_l}$, $v=(2\sqrt{\rho_r} -\sqrt{\rho_l})/\sqrt{\rho_l}$, and the dip density $\rho_{min}=(2\sqrt{\rho_r} -\sqrt{\rho_l})^2$. We emphasize, however, that the equivalence of the leading edge with a gray soliton holds only locally since the DSW is strictly speaking a modulated nonlinear wave.
\begin{figure}[tb]
{\includegraphics[width=4.25cm]{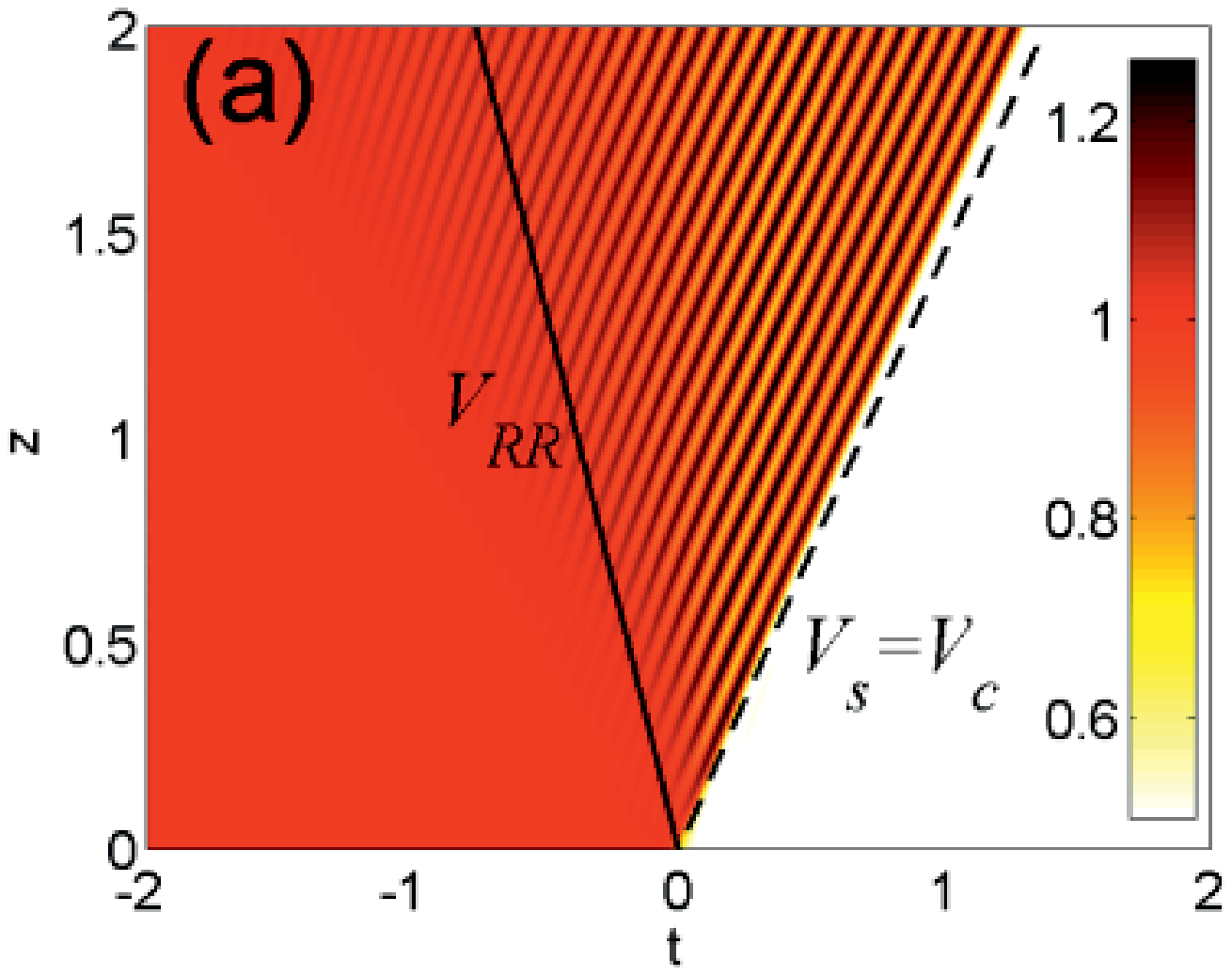}}
{\includegraphics[width=4.25cm]{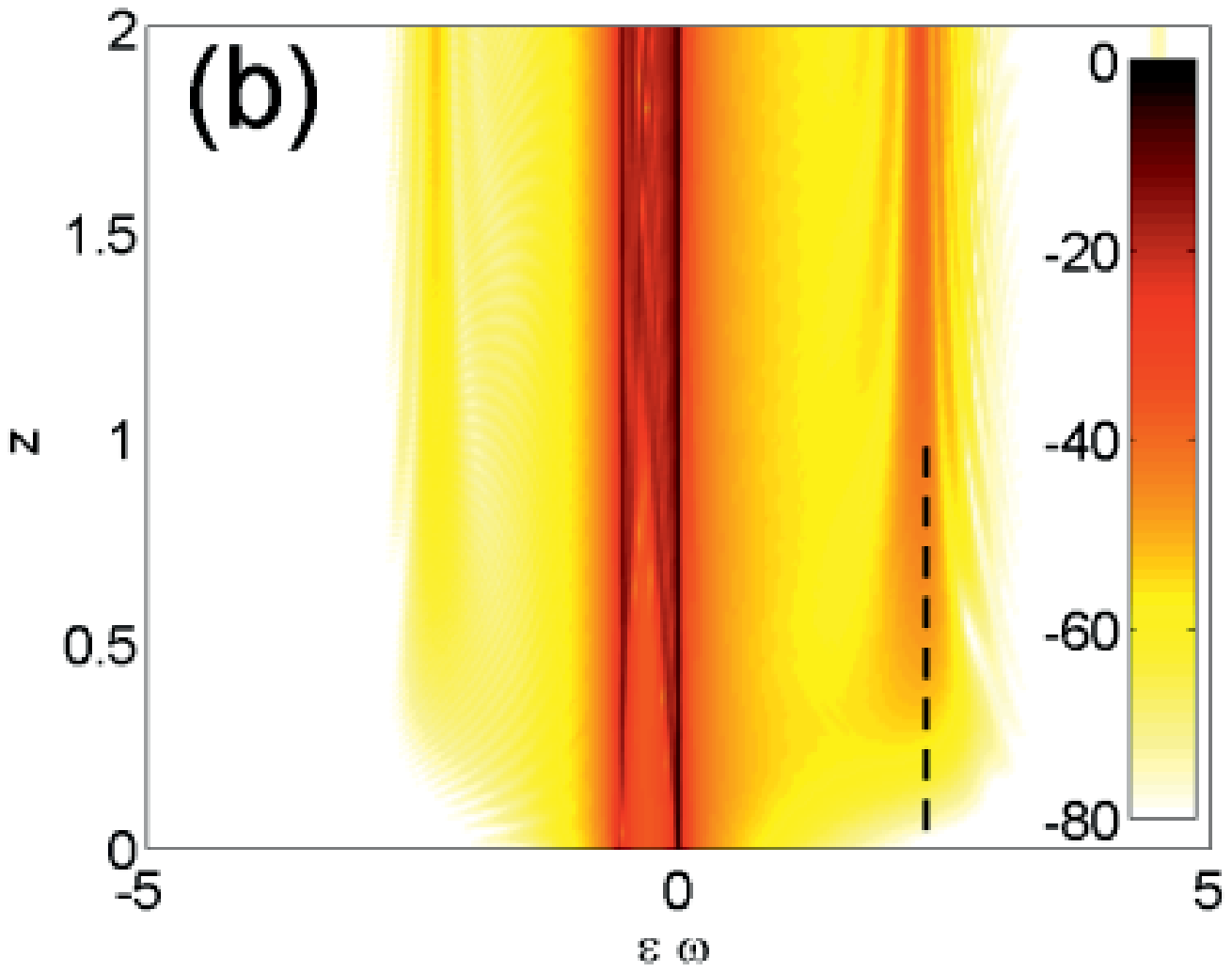}}
{\includegraphics[width=4.25cm]{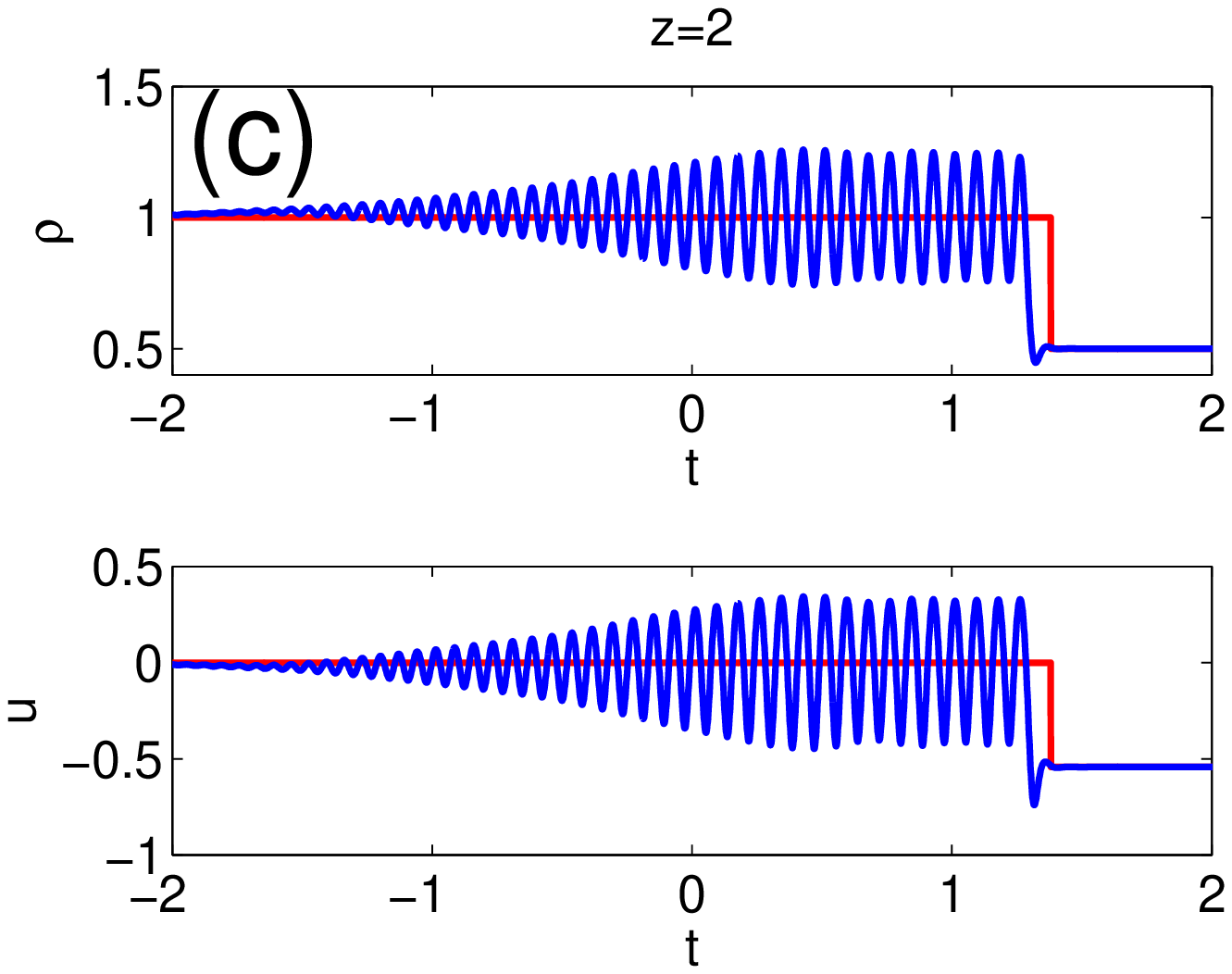}}
{\includegraphics[width=4.25cm]{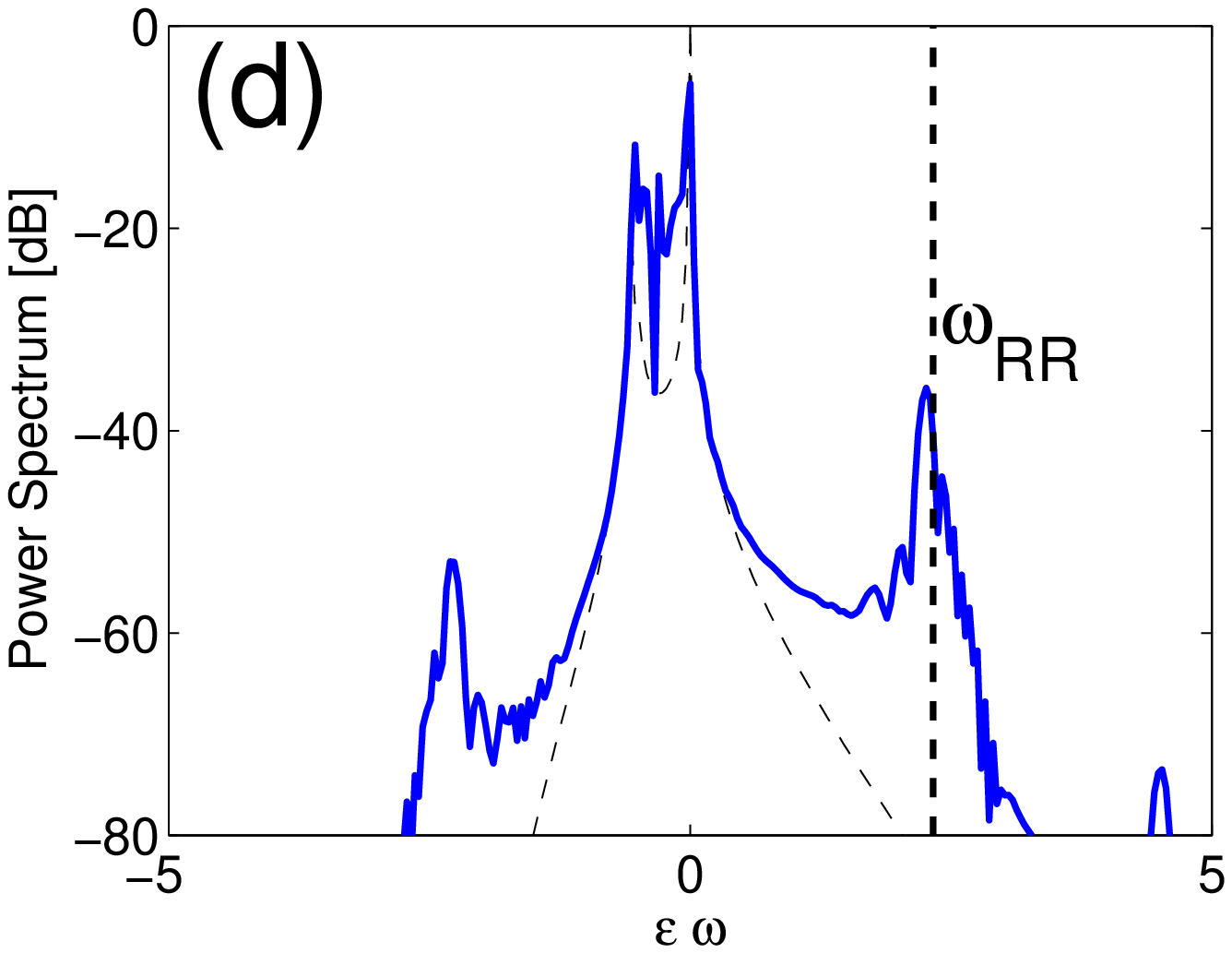}}
\caption{(Color online) 
As in Fig. 1 with larger 3-HOD $\beta_3=-1$. Here $V_c=0.69$ and $u_r=-0.543$ are the parameters of the classical SW [shown in red in (c)] from Eq. (\ref{RHtod}). Solid line in (a) indicates the velocity $V_{RR}$ of the RR.} 
\label{fig2}
\end{figure}
In this regime, if we account for $k_{nl}=k_{nl}^{sol}-k_{nl}^{RR}=-\frac{1}{\varepsilon} \rho_l$ arising from the soliton $k_{nl}^{sol}= \rho_l/\varepsilon$ and the cross-induced contribution $k_{nl}^{RR}=2 \rho_l/\varepsilon$ to the RR,  Eq. (\ref{omRR}) explicitly reads as
\begin{equation} \label{TOD}
\frac{\beta_3}{6}(\varepsilon \omega)^3 + \frac{\beta_2}{2}(\varepsilon \omega)^2 -V_s (\varepsilon \omega) + \rho_l=0.
\end{equation}
Real solutions $\omega=\omega_{RR}$ of Eq. (\ref{TOD}), with $V_s=V_l \equiv 2\sqrt{\rho_r} -\sqrt{\rho_l}$ correctly predicts the RR as long as $|\beta_3| \lesssim 0.5$,
as shown by the  dNLSE simulation in Fig. \ref{fig1}. The DSW displayed in Fig. \ref{fig1}(a) clearly exhibits a spectral RR peak besides spectral shoulders due to the oscillating front, as shown by the spectral evolution in Fig. \ref{fig1}(b) and the output spectrum in Fig. \ref{fig1}(d). Perfect agreement is found between the RR peak obtained in the numerics and the prediction [dashed vertical line in Fig. \ref{fig1}(b,d)] from Eq. (\ref{TOD}) with velocity $V_s=V_l$ characteristic of the integrable limit ($\beta_3=0$, snapshots in Fig. \ref{fig1}c). Indeed, in this regime, the DSW leading edge is nearly unaffected by 3-HOD, whereas using the velocity $V_c$ [Eq. (\ref{RH})] of the equivalent classical SW [reported for comparison in Fig. \ref{fig1}(c)] would miss the correct estimate of $\omega_{RR}$. We also point out that $k_{nl}$ represents a small correction, so $\omega_{RR}$ can be safely approximated by dropping the last term in Eq. (\ref{TOD}) to yield
$\varepsilon \omega_{RR}=\frac{3}{2\beta_3}\left( -\beta_2 \pm \sqrt{\beta_2^2+ 8V_s \beta_3/3} \right)$, or $\varepsilon \omega_{RR}=-\frac{3\beta_2}{\beta_3}$ for $\beta_3 V_s \rightarrow 0$.
\begin{figure}[t]
{\includegraphics[width=4.25cm]{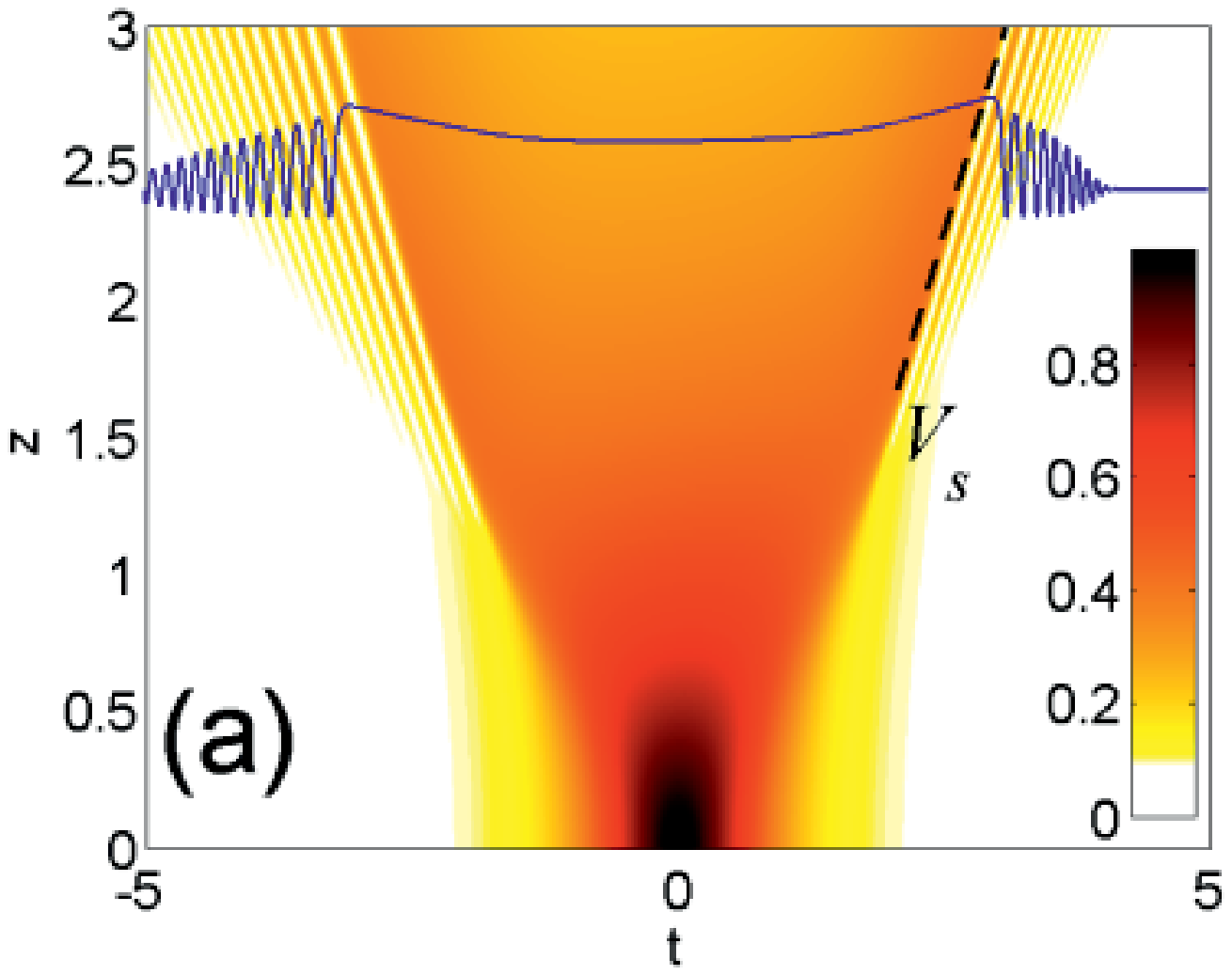}}
{\includegraphics[width=4.25cm]{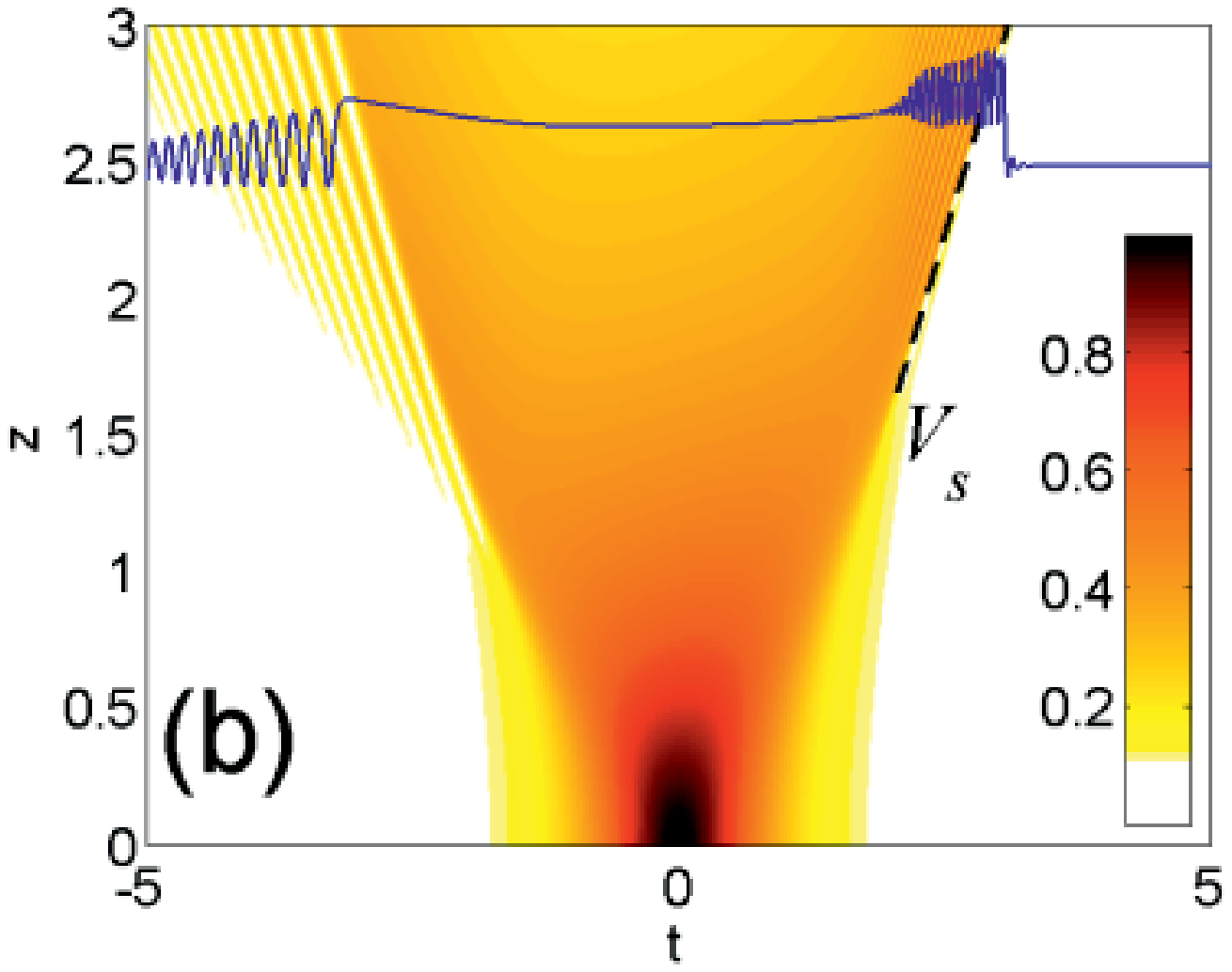}}
{\includegraphics[width=4.25cm]{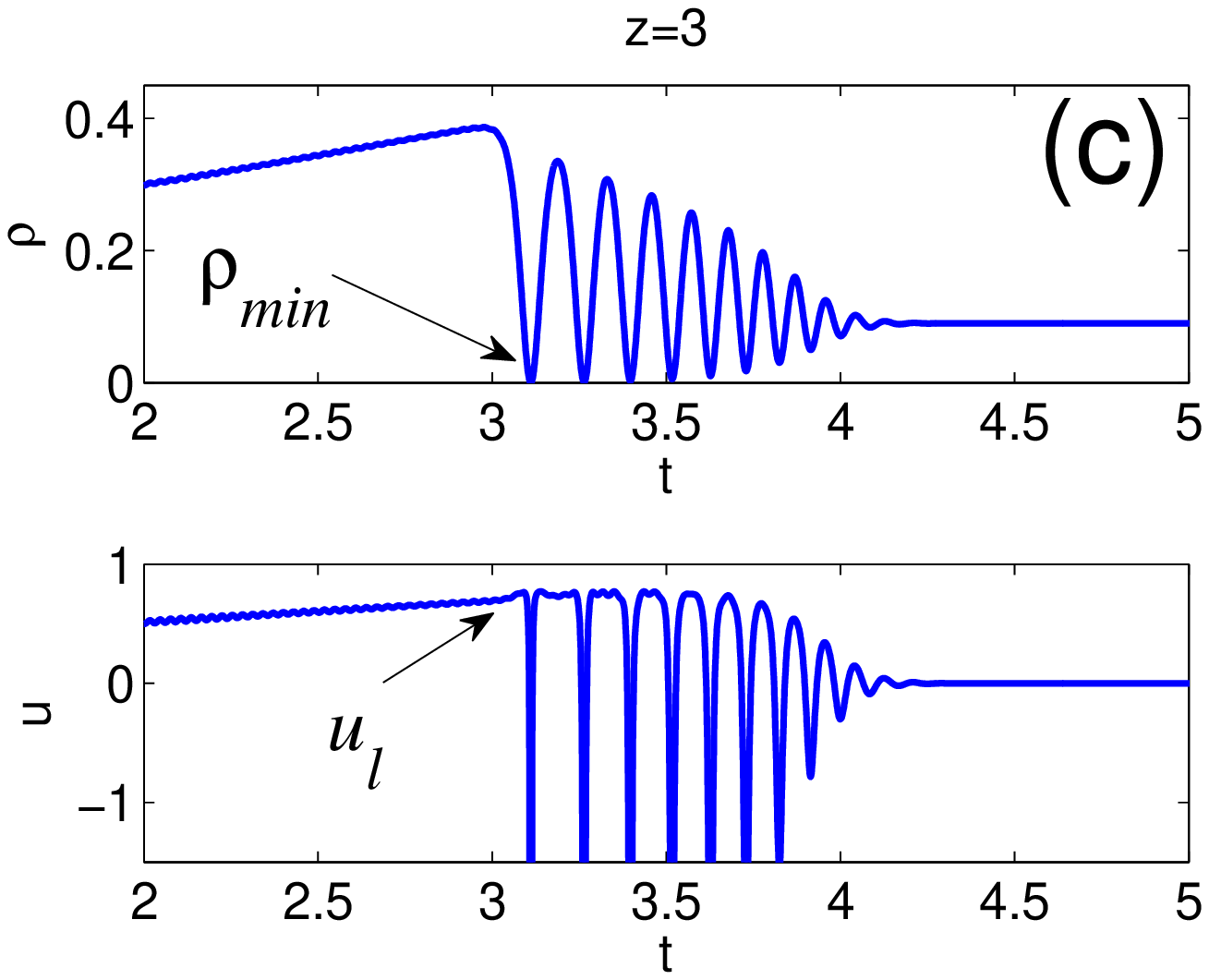}}
{\includegraphics[width=4.25cm]{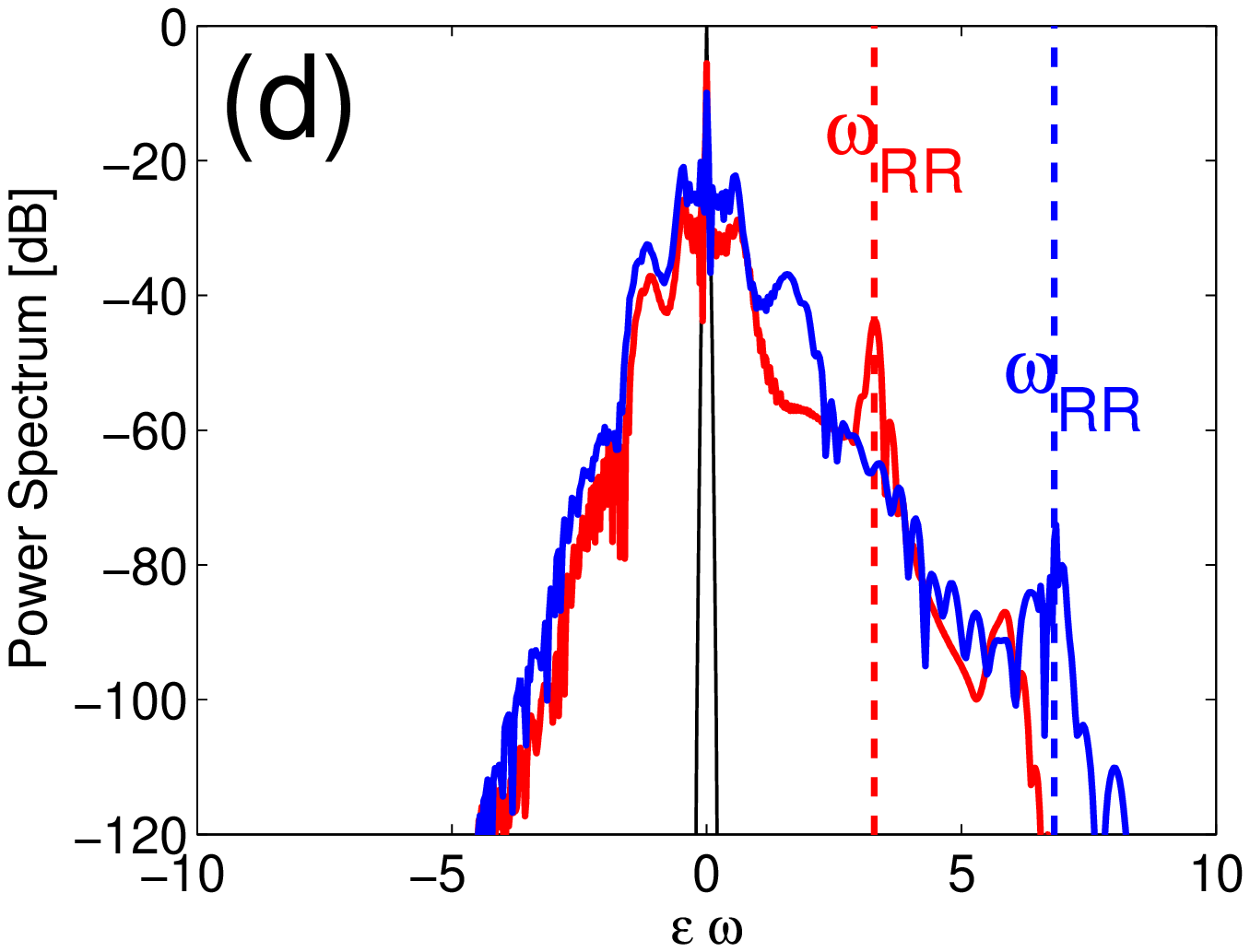}}
\caption{(Color online) 
Radiating DSW from Gaussian pulses with small background $\nu^2=0.09$: (a) $\beta_3=-0.35$; (b) $\beta_3=-0.6$;
(c) Parameters determining the leading edge velocity $V_s=\sqrt{\rho_{min}} + u_l=\sqrt{10^{-4}}+0.76=0.77$ [dashed line in (a)]; 
(d) Output spectra (blue, $\beta_3=-0.35$; red, $\beta_3=-0.6$).
The dashed lines give $\varepsilon \omega_{RR}$ calculated from Eq. (\ref{TOD}).
Here $\varepsilon=0.03$; blue curves in (a),(b) depict output shapshots.} 
\label{fig3}
\end{figure}

When $|\beta_3|$ grows larger, the aperture of the shock fan reduces, until eventually the DSW resembles a single traveling front, i.e. a classical SW \cite{El}.  In this regime, we find that Eq. (\ref{TOD}) still gives the correct frequency $\omega_{RR}$ provided that $V_s$ is taken as the Rankine-Hugoniot velocity $V_c$ of the equivalent classical SW calculated for $\beta_3 \neq 0$ [Eq. (\ref{RHtod})]. This case is illustrated in Fig. \ref{fig2} for $\beta_3=-1$. The RR becomes clearly visible in the temporal evolution [Fig. \ref{fig2}(a,c)], and is sufficiently strong to generate $-\omega_{RR}$ via four-wave mixing [Fig. \ref{fig2}(b-d)]. Perfect agreement between the numerics and the value predicted from Eq. (\ref{TOD}), once we set $V_s=V_c$, is found also in this case. 

The behaviors of step initial data are basically recovered for pulse waveforms that are more manageable in experiments. Figure \ref{fig3} shows the transition from the perturbative [Fig. \ref{fig3}(a)] to the non-perturbative [Fig. \ref{fig3}(b)] regime, for an input gaussian pulse $\psi_0=\nu+(1-\nu) \exp(-t^2)$ with background to peak density ratio $\nu^2=0.09$.
As shown in Fig. \ref{fig3}(a), for relatively small $\beta_3$, two asymmetric DSWs emerge from wave-breaking points on the two pulse edges, which occur at different distances due to broken symmetry in time caused by 3-HOD. Phase-matching is achieved only for the DSW traveling with $V_s>0$. The corresponding $\omega_{RR}$ can be obtained from Eq. (\ref{TOD}) provided we set $V_s=V_l$, with the DSW leading edge velocity being (following the discussion of Fig. \ref{fig1}) $V_l=\sqrt{\rho_{min}} + u_l$, where the minimum density and the correction $u_l$ due to the local non-zero chirp are evaluated numerically after wave-breaking as shown in Fig. \ref{fig3}(c). Also in this case, a larger $|\beta_3|$ results in a narrower fan, until eventually a simple front is left which strongly radiates, as shown in Fig. \ref{fig3}(b). In this regime, a good approximation of the front velocity is obtained by the approximating classical SW in Eq. (\ref{RHtod}). In both the regimes shown in Fig. \ref{fig3}(a) and (b), Eq. (\ref{TOD}) provides a good description of the RR frequency observed in the numerics [see Fig. \ref{fig3}(d)]. Notice also that, for symmetry reasons, sign reversal of 3-HOD (i.e., $\beta_3>0$) simply results into RR with opposite frequency, generated by the DSW with opposite velocity ($V_s<0$, left DSW).

\begin{figure}[!tb]
{\includegraphics[width=4.25cm]{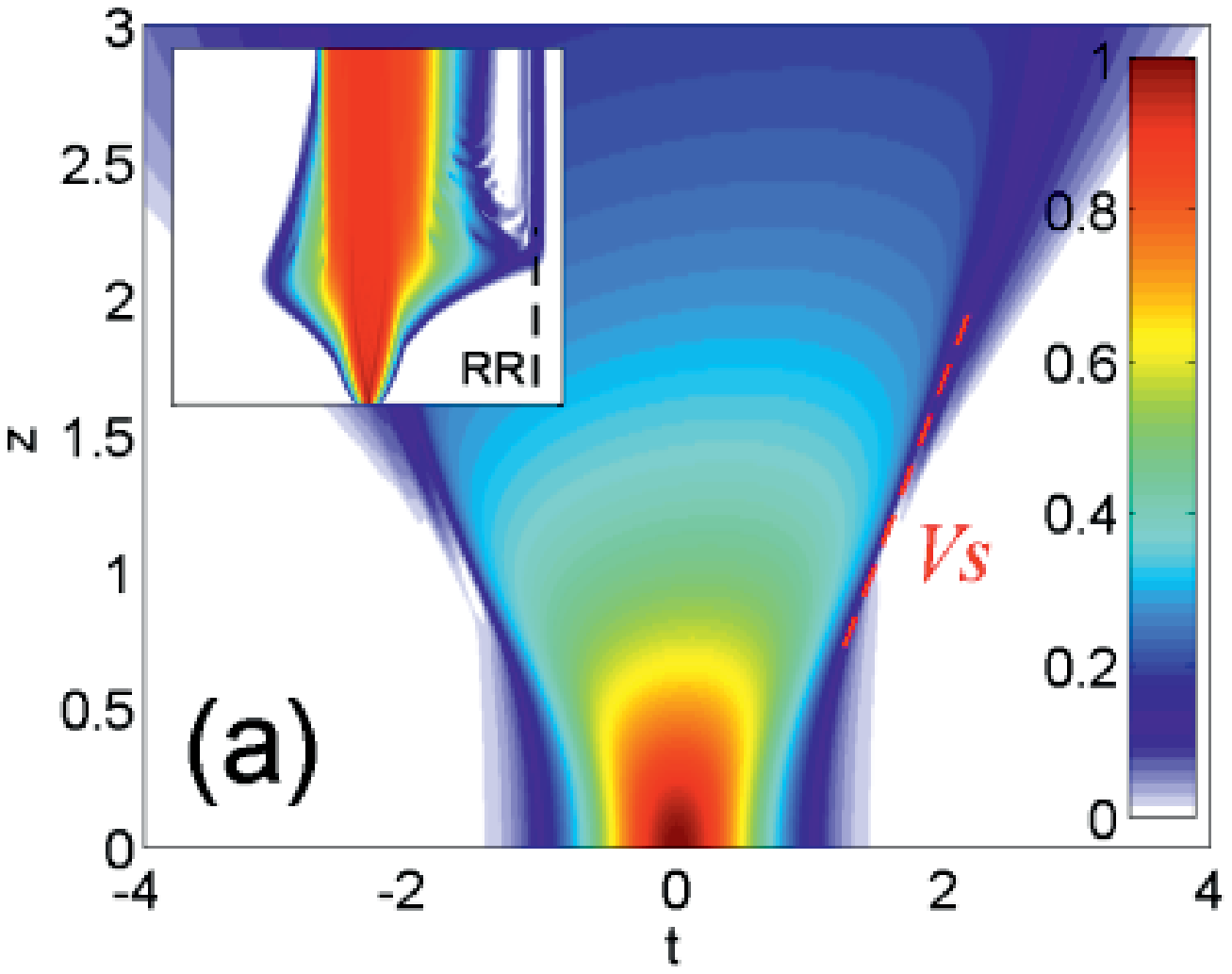}}
{\includegraphics[width=4.25cm]{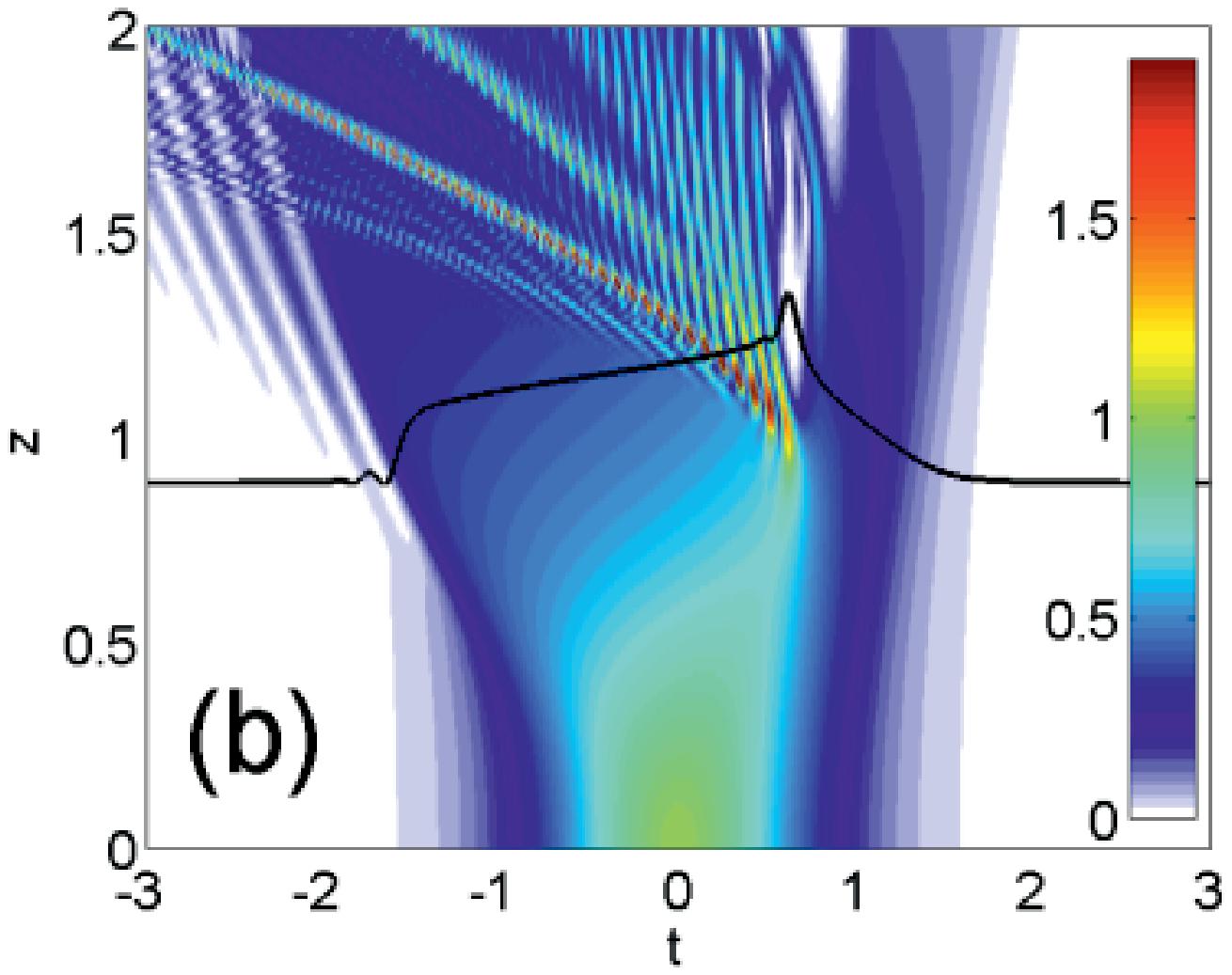}}
\caption{(Color online) 
(a) Temporal evolution of a Gaussian pulse without background, for $\beta_3=-0.35$. Inset shows spectral evolution. (b) Mixed hyperbolic-elliptic type of wave-breaking for large 3-HOD $\beta_3=-2$, $\nu^2=0.01$. Solid line, snapshot at $z=0.9$. Here $\varepsilon=0.03$.} 
\label{fig4}
\end{figure}

We also emphasize two important points: (i) RR occurs also in the limit of vanishing background $\nu=0$, as shown in Fig. \ref{fig4}(a), allowing us to conclude that a bright pulse does not need to be a soliton (as in the fNLSE, $\beta_2=-1$) to radiate. Importantly, experimental evidence for such RR scenario was reported very recently in fiber optics \cite{Webb13}, without explaining the underlying mechanism, which our theory individuates in the shock formation. Indeed, the physical parameters used in Fig. 1 of Ref. \cite{Webb13}, i.e. power $P = 600$ W, pulse duration $T_0 =1$ ps, nonlinear coefficient $\gamma=2.5$ W$^{-1}$ km$^{-1}$, dispersion $\partial^2_\omega k=7.5$ ps$^2$/km, $\partial^3_\omega k=0.2$ ps$^3$/km, gives normalized parameters $\varepsilon \simeq 0.07$ and $\beta_3 \simeq 0.37$, typical of the wave-breaking regime ($\varepsilon \ll 1$) with perturbative 3-HOD. Since $\beta_3>0$, the radiating shock turns out to be the one on the leading edge ($t<0$), and its velocity $V_s = -0.75$, inserted in Eq. (\ref{omRR}), gives a negative [opposite of Fig. \ref{fig4}(a)] frequency detuning  $\Delta f_{RR}=\omega_{RR} T_0^{-1}/2\pi \simeq 13$ THz, in excellent agreement with the value reported in Ref. \cite{Webb13}. 
(ii) a limitation exists (regardless of $\nu$) on the value of $|\beta_3|$ to observe RR, since large 3-HOD feature a qualitative different wave-breaking mechanism, as shown in Fig. \ref{fig4}(b) for $\beta_3=-2$. While the non-radiating (left) DSW simply develop at shorter $z$ without qualitative changes, on the right ($t>0$) the pulse undergoes a different catastrophe, reminiscent of the fNLSE. Indeed in this case the eigenvelocities $V^\pm=\beta_2u+\beta_3 u^2/2 \pm \sqrt{\rho (\beta_2+ \beta_3 u)}$ become locally complex conjugate where $u>0$, implying that Eqs. (\ref{cons1}-\ref{cons2}) show a mixed hyperbolic-elliptic behavior reminiscent of the so-called transsonic flow \cite{transsonic}.

\begin{figure}[!tb]
{\includegraphics[width=4.25cm]{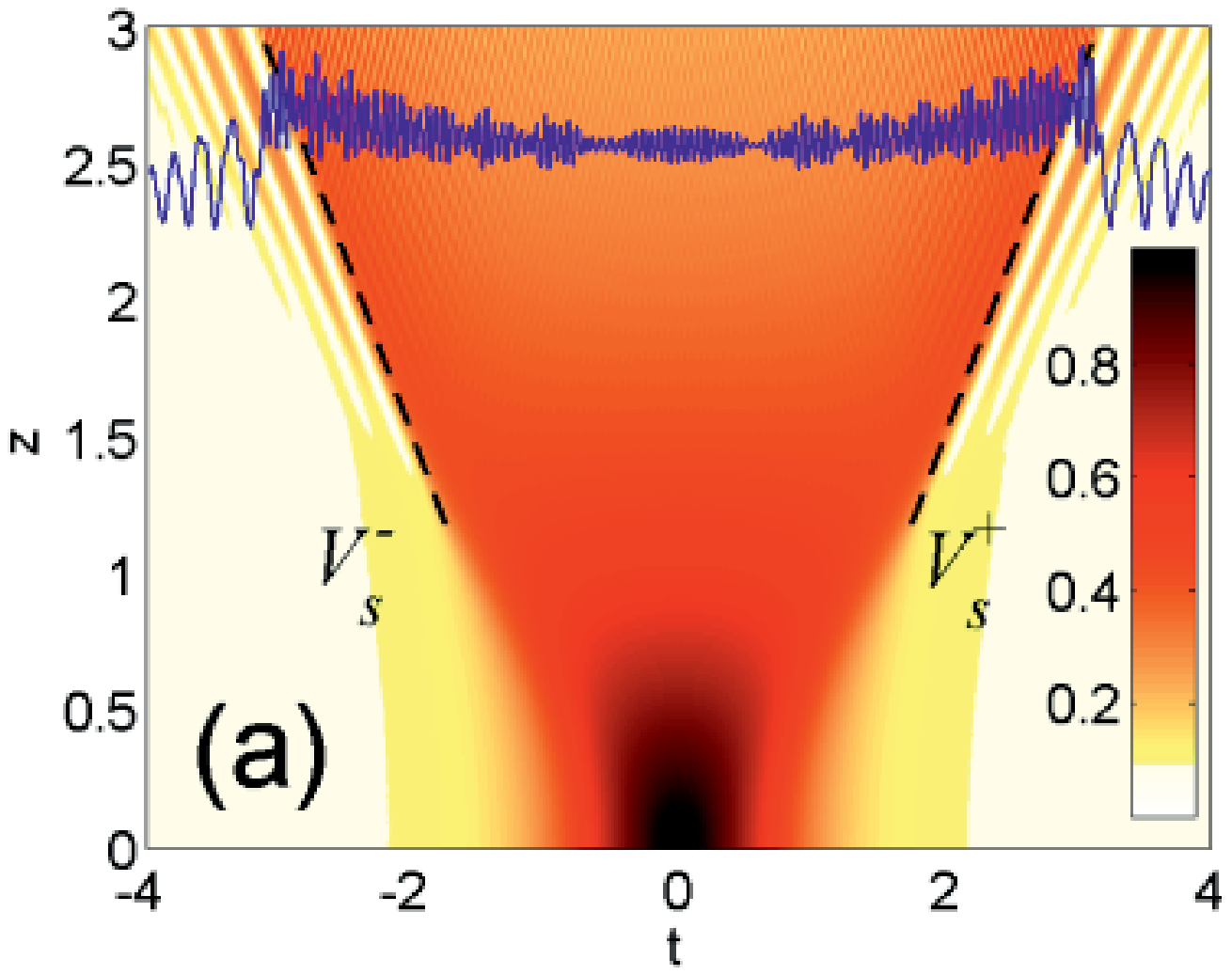}}
{\includegraphics[width=4.25cm]{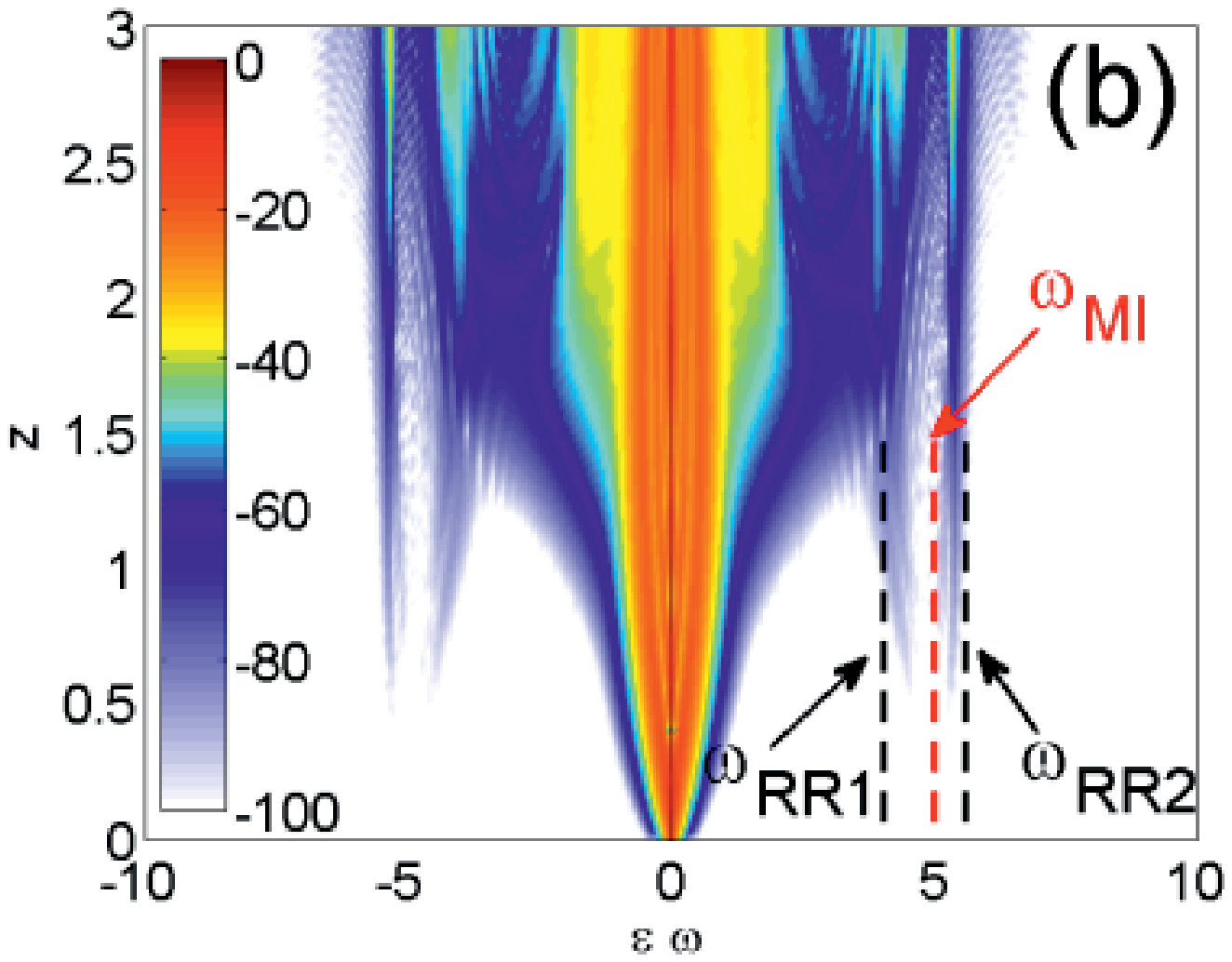}}
\caption{(Color online) RR ruled by 4-HOD ($\beta_3=0$, $\beta_4=-0.5$): 
(a) temporal dynamics (blue curve depicts output snapshot); (b) spectral evolution, black dashed lines are prediction from Eq. (\ref{omRR}), red dashed line is MI peak gain. 
Here $\varepsilon=0.05$, input Gaussian pulse with background $\nu^2=0.09$.} 
\label{fig5}
\end{figure}

A  completely different scenario occurs when the dispersive correction is due to 4-HOD. In this case, the shock formation can compete with a different breaking mechanism, namely modulational instability (MI). The latter, characteristic of the fNLSE, is known to extend to the defocusing regime $\beta_2=1$, whenever $\beta_4<0$ \cite{MI} (see also supplemental material \cite{SM}). Moreover, the problem is symmetric in time and the shocks from both edges of a pulse radiate. Overall four RR frequencies result from Eq. (\ref{omRR}), which is now fourth-order: the frequency pair $\omega_{RR1},-\omega_{RR2}$ ($\omega_{RR1,RR2}>0$) and the opposite pair $-\omega_{RR1}, \omega_{RR2}$, induced by the shock with positive ($V_s>0$) and negative velocity ($V_s>0$), respectively. Since the MI has a spectral narrow bandwidth that turns out to lie exactly in between the two frequencies $\omega_{RR1,RR2}$  (arising from shocks on opposite edges), it serves as a seed for the RR. This is shown in Fig.~\ref{fig5}: the pair of twin-band RR starts to grow, triggered by MI, even during the process of pulse edge steepening ($z < 1$), while becoming prominent as the DSWs form and travel with definite velocities (here $V_s=\pm 0.77$). The four-band RR from the spectral evolution  in Fig.~\ref{fig5}(b) fits well the prediction from Eq. (\ref{omRR}) (dashed lines), while coexistence of the two wave-breaking phenomena (MI and DSW) is clearly visible in the output snapshot in Fig.~\ref{fig5}(a).


In summary, we have demonstrated that higher-order dispersive corrections force DSWs to radiate, 
following different observable cross-over scenarios depending on the nature and magnitude of such corrections.\\
Funding from MIUR (grant PRIN  2009P3K72Z)  is gratefully acknowledged.


\begin{thebibliography}{}

\bibitem{collisionless}
R. Z. Sagdeev, 
Sov. Phys. Tech. Phys. {\bf 6}, 867 (1962); 
R.J. Taylor, D.R. Baker, and H. Ikezi, Phys. Rev. Lett. {\bf 24}, 206 (1970);
A.V. Gurevich and L.P. Pitaevskii, 
Sov. Phys. JETP {\bf 38}, 291 (1974);

\bibitem{water1} 
T.B. Benjamin and M.J. Lighthill,
Proc. Roy. Soc. Lond. A, {\bf 224}, 448 (1954);
D. H. Peregrine,  
J. Fluid Mech. {\bf 25}, 321 (1966).

\bibitem{BEC}
Z. Dutton {\it et al.,} 
Science {\bf 293}, 663 (2001); 
A.M. Kamchatnov, A. Gammal, and R.A. Kraenkel, 
Phys. Rev. A {\bf 69}, 063605 (2004);
R. Meppelink {\it et al.,}  
\pra {\bf 80}, 043606 (2009).

\bibitem{Hoefer06} 
M. A. Hoefer, M. J. Ablowitz, I. Coddington, E. A. Cornell, P. Engels, and V. Schweikhard,
\pra {\bf 74}, 023623 (2006); J.J. Chang, P. Engels, and M.A. Hoefer, 
Phys. Rev. Lett. {\bf 101}, 170404 (2008).

\bibitem{temporal} J. E. Rothenberg and D. Grischkowsky, 
Phys. Rev. Lett. {\bf 62}, 531 (1989);
Y. Kodama, S. Wabnitz, and K. Tanaka,
Opt. Lett. {\bf 21}, 719 (1996);
C. Conti, S. Stark, P. St. J. Russell, and F. Biancalana, 
Phys. Rev. A {\bf 82,} 013838 (2010);
M. Conforti, F. Baronio, and S. Trillo,
Opt. Lett. {\bf 37,} 1082 (2012);
J. Fatome {et al.}, 
{\it Observation of colliding optical undular bores spontaneously generated via four-wave mixing in optical fibres},
submitted to Nature Photonics.

\bibitem{spatial}
W. Wan, S. Jia, and J. W. Fleischer, 
Nature Phys. {\bf 3}, 46 (2007);
Phys. Rev. Lett. {\bf 99}, 223901 (2007);
N. Ghofraniha, C. Conti, G. Ruocco, and S. Trillo, 
Phys. Rev. Lett. {\bf 99}, 043903 (2007);
C. Conti {\it et al.,} 
Phys. Rev. Lett. {\bf 102}, 083902 (2009).

\bibitem{water} 
N.F. Smyth, P.E. Holloway, 
J. Phys. Oceanogr. {\bf 18}, 947(1988);
G.~A. El, H.~J. Grimshaw, A.~M. Kamchatnov, 
Stud. Appl. Math. {\bf 114}, 395 (2005);
J. R. Apel, 
J. Phys. Oceanogr. {\bf 33}, 2247 (2003).

\bibitem{quantum}
E. Bettelheim, A.G. Abanov, P. Wiegmann,
Phys. Rev. Lett. {\bf 97}, 246401 (2006).
 
\bibitem{electrons}
Y. C. Mo  {\it et al.}, 
Phys. Rev. Lett. {\bf 110}, 084802 (2013).

\bibitem{magma}
N. K. Lowman and M. A. Hoefer, 
J. Fluid Mech. {\bf 718}, 524 (2013).

 \bibitem{granular}
P. Lorenzoni and S. Paleari, 
Phys. D {\bf 221}, 110 (2006);
A. Molinari and C. Daraio, 
Phys. Rev. E {\bf 80}, 056602 (2009).

\bibitem{disorder}
N. Ghofraniha, S. Gentilini, V. Folli, E. Del Re, and C. Conti,
Phys. Rev. Lett. {\bf 109}, 243902 (2012).

\bibitem{BO} 
P. D. Miller and Z. Xu,  
Commun. Pure Appl. Math.  {\bf 64}, 205 (2010).

\bibitem{dNLSE} 
A.V. Gurevich and A. L. Krylov, 
Sov. Phys. JETP {\bf 65}, 944 (1987);
A.V. Gurevich, A. L. Krylov, and G.A. El,
Sov. Phys. JETP {\bf 74}, 957 (1992).

\bibitem{RRoptics}
P. K. A. Wai, C. R. Menyuk, Y. C. Lee, and H. H. Chen,
Opt. Lett. {\bf 11}, 464 (1986); {\em ibidem}, {\bf 12}, 628 (1987);
P. K. A. Wai, H. H. Chen, and Y. C. Lee, 
Phys. Rev. A {\bf 41}, 426 (1990).

\bibitem{AK95}
N. Akhmediev and M. Karlsson, 
Phys. Rev. A {\bf 51}, 2602 (1995).

\bibitem{RRothers}
H. H. Kuehl and C. Y. Zhang, 
Phys. Fluids B {\bf 2}, 889 (1990); 
V. I. Karpman and H. Schamel, 
Phys. Plasmas {\bf 4}, 120 (1997);
V. I. Karpman, 
Phys. Rev. E {\bf 58}, 5070 (1998).

\bibitem{RRdark} 
V. V. Afanasjev and Y. S. Kivshar, and C. R. Menyuk,
Opt. Lett. {\bf 21}, 1975 (1996);
C. Milian, D. V. Skryabin, and A. Ferrando, 
Opt. Lett. {\bf 34}, 2096 (2009).




\bibitem{SGRMP10}
D. V. Skryabin and A. V. Gorbach,
Rev. Mod. Phys. {\bf 82}, 1287 (2010).

\bibitem{DudleyRMP06}
J. M. Dudley, G. Genty,  and S. Coen, 
\rmp~{\bf 78}, 1135 (2006).

\bibitem{SChydro}
A. Chabchoub, N. Hoffmann, M. Onorato, G. Genty, J. M. Dudley, and N. Akhmediev, 
Phys. Rev. Lett. {\bf 111}, 054104 (2013).



\bibitem{Stark11}
S. P. Stark, A. Podlipensky, and P. St. J. Russell,
Phys. Rev. Lett. {\bf 106}, 083903 (2011).

\bibitem{Joly11}
N.Y. Joly, J. Nold, W. Chang, P. H\"olzer, A. Nazarkin, G. K. L. Wong, F. Biancalana, and P. St. J. Russell,
Phys. Rev. Lett. {\bf 106}, 203901 (2011);
M. F. Saleh {\it et al.}, 
Phys. Rev. Lett. {\bf 107}, 203902 (2011).

\bibitem{Erkintalo12}
M. Erkintalo, Y.Q. Xu, S.G. Murdoch, J.M. Dudley, and G. Genty,
Phys. Rev. Lett. {\bf 109}, 223904 (2012).


\bibitem{Colman12}
P. Colman {\it et al.}, 
Phys. Rev. Lett. {\bf 109}, 093901 (2012).

 
\bibitem{Bache10}
B. B. Zhou, A. Chong, F. W. Wise, and M. Bache, 
Phys. Rev. Lett. {\bf 19}, 18754 (2012);
M. Conforti and F. Baronio,
J. Opt. Soc. Am. B {\bf 30,} 1041 (2013).

\bibitem{rubino12}
E. Rubino, J. McLenaghan, S. C. Kehr, F. Belgiorno, D. Townsend, S. Rohr, C. E. Kuklevicz, U. Leonhardt, F. K\"onig, and D. Faccio,
Phys. Rev. Lett. {\bf 108}, 253901 (2012).

\bibitem{spatialHOD} 
HOD-terms are relevant also for spatial problems to account for non-paraxial effects,
see e.g. C. Conti, G. Ruocco, and S. Trillo, Phys. Rev. Lett. {\bf 95}, 183902 (2005). 

\bibitem{SC12} 
Y. Liu, H. Tu, and S. A. Boppart
Opt. Lett. {\bf 37}, 2172 (2012)

\bibitem{Webb13} 
K. E. Webb, Y. Q. Xu, M. Erkintalo, and S. G. Murdoch,
Opt. Lett. {\bf 38}, 151 (2013).

\bibitem{SM} See Supplemental Material at [URL will be inserted by publisher] for more technical details on the normalization of Eq. (1),
evaluation of Rankine-Hugoniot velocity, and MI analysis in the presence of 4-HOD.

\bibitem{nota_wp}
This simply means, without loss of generality, that in real world units $\omega_p$  coincides with $\omega_0$, around which $d(\partial_t)$ in Eq. (\ref{wdnls}) is expanded

\bibitem{notaRR}
The nonlinear contribution to the wawenumber of the resonant radiation is induced by cross-phase modulation with a non-zero background, on top of which RR propagates.

\bibitem{Whitham74}
G. B. Whitham, {\it Linear and Nonlinear Waves} (Wiley, New York, 1974).

\bibitem{El}
in the intermediate regime $0.5<|\beta_3|<1$, an estimate of the leading edge velocity could be given via a non-integrable formulation of the Whitham modulation theory,
see G. A. El, Chaos {\bf 15}, 1 (2005) and Ref. [10], which we will develop elsewhere.

\bibitem{transsonic} 
J.C. Di Franco, P. D. Miller, and B. K. Muite, 
Acta Math. Scientia {\bf 31B}, 2343 (2011).


\bibitem{MI} 
S. B. Cavalcanti, J. C. Cressoni, H. R. da Cruz, and A. S. Gouveia-Neto,
\pra {\bf 43}, 6162 (1991);
S. Pitois and G. Millot, 
Opt. Commun. {\bf 226} 415 (2003);
J. D. Harvey, R. Leonhardt, S. Coen, G. K. L. Wong, J. C. Knight, W. J. Wadsworth, and P. S. J. Russell,
Opt. Lett. {\bf 28}, 2225 (2003).



\end{thebibliography}
\end{document}